\documentclass[conference]{IEEEtran}
\IEEEoverridecommandlockouts
\usepackage{cite}
\usepackage{amsmath,amssymb,amsfonts}
\usepackage{graphicx}
\usepackage{textcomp}
\usepackage{xcolor}
\usepackage{multirow}
\usepackage{hyperref}
\usepackage{booktabs}
\usepackage{caption}
\usepackage{subcaption}
\usepackage{balance}
\usepackage[noend]{algpseudocode}

\def\BibTeX{{\rm B\kern-.05em{\sc i\kern-.025em b}\kern-.08em
    T\kern-.1667em\lower.7ex\hbox{E}\kern-.125emX}}
\begin{document}

\title{Debt-Financed Collateral and Stability Risks in the DeFi Ecosystem}

\author{\IEEEauthorblockN{Michael Darlin}
\IEEEauthorblockA{\textit{School of Management and} \\
\textit{Yale Institute for Network Science}\\
Yale University, USA}
\and
\IEEEauthorblockN{Georgios Palaiokrassas}
\IEEEauthorblockA{\textit{Department of Electrical Engineering and} \\
\textit{Yale Institute for Network Science}\\
Yale University, USA}
\and
\IEEEauthorblockN{Leandros Tassiulas}
\IEEEauthorblockA{\textit{Department of Electrical Engineering and} \\
\textit{Yale Institute for Network Science}\\
Yale University, USA}
}

\maketitle

\begin{abstract}
The rise of Decentralized Finance (``DeFi") on the Ethereum blockchain has enabled the creation of lending platforms, which serve as marketplaces to lend and borrow digital currencies. Initially, we categorize the activity of lending platforms within a standard regulatory framework. We then propose an Ethereum address grouping algorithm using activity over  DeFi protocols and employ a novel classification algorithm to calculate the percentage of fund flows into DeFi lending platforms that can be attributed to debt created elsewhere in the system (``debt-financed collateral''). Based on our results, we conclude that the wide-spread use of stablecoins as debt-financed collateral increases financial stability risks in the DeFi ecosystem.
\end{abstract}

\begin{IEEEkeywords}
DeFi, dept-financed collateral, address clustering, heuristics, financial stability risks
\end{IEEEkeywords}

\section{Introduction}
DeFi applications have seen a rapid increase in popularity and usage since the beginning of 2019. One of the core functionalities of DeFi is the creation of debt through collateralized lending. 
In this paper, we examine whether debt financing was the main source of funds deposited into DeFi protocols, and their risks to overall financial stability. We then develop an algorithm to track debt and collateral balances across user addresses and calculate the proporiton of ``debt-financed collateral'' (DFC)  into DeFi applications. 

We conclude that DFC has been used extensively across DeFi protocols. While debt usage in general increases financial stability risks, DFC in particular raises risks because the method links debt across multiple protocols. This interconnectedness increases the risk of contagion in the case of  a run on lending platform deposits, or in the event of a lending platform's operational failure. 

This paper contributes the following insights:
\begin{enumerate}
    \item \textbf{Regulatory assessment of DeFi risks}: We apply the ``economic functions" framework used by macroprudential regulators to classify and assess the financial risks posed by DeFi lending platforms.
    \item \textbf{Address Clustering Algorithm}: To the best of our knowledge, we are the first to propose an algorithm that clusters Ethereum addresses using the activity and interactions of users in the DeFi ecosystem.
    \item \textbf{Debt-creation algorithm}: We devise a novel algorithm to group addresses, classify Ethereum transactions, and calculate the percentage of DFC.
    \item \textbf{System-level aggregate analysis}: We apply the algorithm to transactions from four major DeFi protocols, and we use the results to assess the risk levels endogenous to DeFi over time, while we release the collected data and our source code for further research\footnote{\url{https://github.com/michael-darlin/debt-financed-collateral}.}.
\end{enumerate}

In Section \ref{sec:DeFi}, we provide an overview of the DeFi ecosystem and of lending platforms specifically. In Section \ref{sec:regAssess}, we provide a risk assessment of lending platforms according to a financial regulatory framework. Our analysis methodology is defined in Section \ref{sec:methodology}, and the algorithm parameters are specified in Section \ref{sec:algorithm}. Section \ref{sec:analysis} details the results of our analysis, and Section \ref{sec:conclusion} concludes with brief remarks on potential avenues for future research. 

\section{Decentralized Finance}
\label{sec:DeFi}
DeFi refers to a set of financial applications operating primarily on top of Ethereum smart contracts.
but also includes a desire to replace existing financial intermediaries with automated financial software \cite{ftMovement}, maintaining an open and transparent financial system, and a commitment to community-led, decentralized governance of applications \cite{cbBeginnersGuide}. 

The move towards a decentralized financial system was recently envisioned by the Basel Committee for Banking Supervision (``BCBS''). In a hypothetical scenario of mostly disintermediated banks, the BCBS concluded customers would have ``a more direct say in choosing the services and the provider rather than sourcing such services via an intermediary bank.'' At the same time, customers would also ``assume more direct responsibility in transactions, increasing the risks they are exposed to'' \cite{soundPractices}. DeFi applications exemplify the possibilities and pitfalls of a ``disintermediated bank'' scenario.
The public permissionless nature of the Ethereum blockchain allows any individual to review, develop or interface with DeFi protocols \cite{cbBeginnersGuide},  which interact with each other, creating a tightly-connected system programmatically and financially \cite{defiCrisis}. 

Although address clustering methods have been proposed for Bitcoin's UTXO model, a limited number of methods have been proposed for Ethereum's account model \cite{victor2020address, wu2022tutela, beres2021blockchain}, focusing in certain airdrops, ICOs or other on-chain activities and interactions. However, the participation of users in DeFi platforms has not yet been exploited towards this direction.

DeFi Total Value Locked (“TVL”) was measured at \$0.9B on April 30, 2020 and on May 17, 2022, it was measured at \$56.9B \cite{defiPulse}. This increase may be attributed to an increased usage of DeFi applications, an increasing leveraged capital, and a more than ten-fold increase in Ethereum prices over the same period. DeFi transactions commonly involve stablecoins, cryptocurrencies such as Tether, USD Coin, and DAI \cite{wsjStablecoins}.

Lending platforms, which are a popular implementation of DeFi technology, allow users to both lend and borrow currency, benefiting from the existence of many ERC-20 tokens \cite{erc20Standard}. Lenders can deposit currency into a lending platform as liquidity for loans, thereby earning interest on their deposits. Borrowers deposit currency as collateral for their own loans, subsequently drawing down debt in a new currency. In essence, funds deposited for lending act as liquidity for funds being requested for borrowing. The largest lending platform, Maker, allows users to borrow only in the Dai stablecoin, which is created by the loan itself and is backed by the loan collateral. Other popular platforms, such as Aave and Compound, allow users to borrow multiple currencies, the liquidity for which is provided by the pooled currencies of other protocol users.

Interest rates for each currency are adjusted, either manually or algorithmically, based on the supply and demand. Debt withdrawn from lending platforms must stay below a certain percentage of the collateral deposited, in order to avoid liquidation. While the collateral stays locked in the smart contract, the debt withdrawn can be used freely, and is essentially new money created through the lending platform.  

\begin{figure}[htb]
    \resizebox{\columnwidth}{!}{%
        \includegraphics{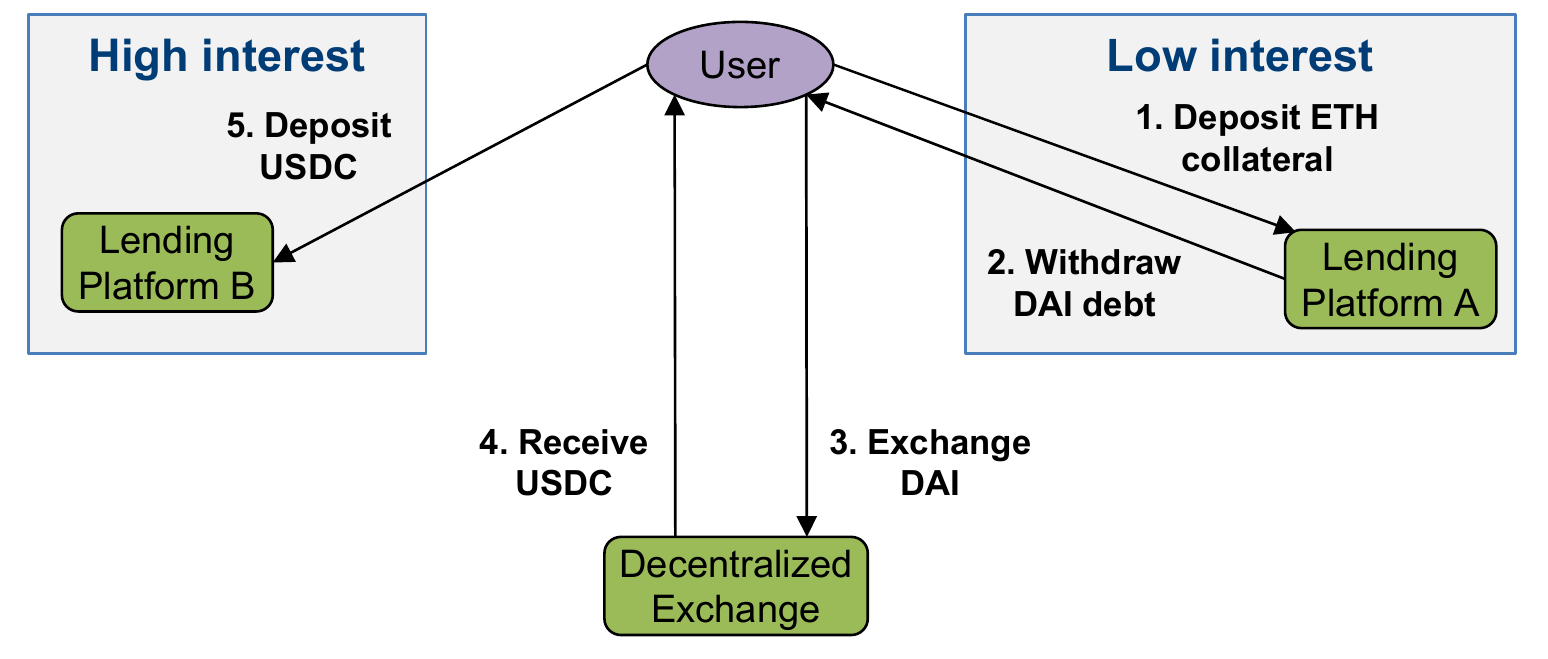}
    }
    \caption{Example transaction sequence involving lending platforms and decentralized exchanges}
    \label{fig:lendPlatformTransac}
\end{figure}

Figure \ref{fig:lendPlatformTransac} illustrates the process of interacting with multiple lending platforms, first to withdraw a loan (Steps 1-2) and then to deposit funds as liquidity for loans (Step 5). In between, users may interact with decentralized exchanges, which allow them to exchange between currencies (Steps 3-4). These transactions may be undertaken to profit from differences in interest rates across currencies and protocols, or to gain exposure to a currency expected to appreciate in price. Under the parameters of Algorithm \ref{algo:classify}, described further in Section \ref{sec:algorithm}, we would classify Steps 1 and 5 of Figure \ref{fig:lendPlatformTransac} as ``Collateral Deposit'' transactions, Step 2 as a ``Debt Create'' transaction, and Steps 3 and 4 as a ``Swap'' transaction. Two additional transaction types in Algorithm \ref{algo:classify}, not depicted in Figure \ref{fig:lendPlatformTransac} for simplicity, are ``Collateral Withdraw'' (the obverse of Step 1) and ``Debt Repay'' (the obverse of Step 2).

In our analysis, we focus on the activities of the largest lending platforms (Aave, Compound and Maker) and one of the most popular decentralized exchanges (Uniswap) \cite{defiPulse}, as these are the drivers of debt creation in DeFi ecosystem.

\section{Regulatory Assessment}
\label{sec:regAssess}
The Financial Stability Board (``FSB"), an international financial monitoring body, defines the non-bank financial intermediation (``NBFI") sector as ``all financial institutions that are not central banks, banks or public financial institutions" \cite{nbfi2020Report}. The NBFI sector has been a focal point for banking regulators since the financial crisis of 2007-09, when asset price bubbles were attributed, in part, to the unregulated activities of non-bank financial institutions \cite{shadowBanking}. 
Recently, the FSB has issued reports on several issues related to cryptocurrencies: global cryptocurrency regulation \cite{cryptoAssets}, disintermediation of financial services through new technologies (including, but not limited to, blockchain) \cite{decentralisedFinance}, and a proposed regulatory framework for stablecoins \cite{stablecoins}. We build on those reports by examining how the activities of specific DeFi protocols (lending pools) could be placed within the FSB's monitoring framework. The FSB concentrates its monitoring function on NBFI activities that increase financial stability risk, analyzed below (Sect. \ref{sec:activities}).

\subsection{Activities-based assessment}
\label{sec:activities}
Regulatory arbitrage (the first of two major NBFI activities covered by the FSB) is defined as activities that mimic the functions of traditional banks (e.g. providing credit to consumers) but are conducted outside existing regulatory areas. These activities enjoy freedom from regulatory purview (hence the term ``arbitrage") but still increase the amount of risk within the financial system \cite{policyFramework}, with lending platforms operating outside the current scope of macroprudential regulations. 

Credit intermediation (the second major NBFI activity) refers to an intermediary assisting in credit creation. Lending platforms fall into this category, as they provide a marketplace for users to lend and borrow funds and we consider which specific types of intermediation \cite{nbfi2020Report} lending platforms provide: 

\emph{i) Maturity transformation} is defined as ``the use of short-term deposits to fund long-term loans" \cite{shadowBanking}. This transformation arises when banks originate loans which are repaid on a long-term schedule, while also holding deposits which may be redeemed at any time. In lending pools, borrowers are under no obligation to repay their loans via a fixed schedule, and they may keep their loan principal for as long as their collateral balance meets the protocol's parameters, while lenders may redeem their deposits at any time. Consequently, lending platforms engage in maturity transformation. 

\emph{ii) Liquidity transformation} is defined as ``the use of liquid instruments to fund illiquid assets" \cite{shadowBanking} and is not present in lending platforms, since both deposits and loans are made in highly liquid currencies.
In theory, lending platforms could accept deposits in liquid currency and provide loans in an illiquid currency, which would qualify as liquidity transformation. However, while smaller currencies may have little liquidity, the majority of DeFi lending and borrowing is concentrated in a few major currencies (Table \ref{tab:coverage}). 
Even when liquidity decreases during a market crash, the liquidity of a crypto-asset should exceed that of an illiquid long-term investment.

\emph{iii) Leverage:} defined as the amount of debt taken on by an individual or entity, commonly measured through the leverage ratio
\cite{leverageDynamics}. 
Since lending platforms allow users to borrow funds, the protocols are engaged in providing leverage.

\emph{iv) Imperfect credit risk transfer} refers to credit risk \cite{gseCRT} not being appropriately priced. This leaves the intermediary exposed to residual credit risk, which does not apply to the activities of lending platforms, since they do not formally engage as intermediaries in credit risk transfer.

\subsection{Economic function classification}
From our activities analysis, we conclude that lending platforms engage in leverage creation and maturity transformation. Using these activities as a filter, we can place lending platforms within the FSB's economic function framework. The FSB defines five economic functions (``EFs"), which may increase the risk to macro-level financial stability \cite{nbfi2020Report}.

\emph{EF1:} Like investment funds (``collective investment vehicles''), lending platforms accept deposits from investors, who receive a return on their deposits through interest and may withdraw their funds at any time. Unlike investment funds, however, lending platforms do not actively invest the funds deposited; interest is earned based on demand from borrowers interacting with the lending platform's smart contracts. As a result, we do not classify lending platforms under EF1.

\begin{table}[htbp]
    \caption{Economic functions}
        \begin{tabular}{@{} p{0.3cm} p{8cm} @{}}
            \hline
            \bf EF & \bf Definition \\ 
            \hline
            EF1 & Management of collective investment vehicles with features that make them susceptible to runs \\
            EF2 & Loan provision that is dependent on short-term funding \\
            EF3 & Intermediation of market activities that is dependent on short-term funding or on secured funding of client assets \\
            EF4 & Facilitation of credit creation \\
            EF5 & Securitisation-based credit intermediation and funding of financial entities \\
            \hline
        \end{tabular}
\end{table}

\emph{EF2:} The FSB defines an example EF2 entity as one that takes ``deposits from retail and wholesale customers that are redeemable at notice or within a short timeframe" \cite{policyFramework}. We have already observed that the main purpose of DeFi protocols is the provision of credit to borrowers. In addition, deposits into DeFi protocol are ``redeemable at notice," providing that a user has no loans outstanding. So, lending platforms can be classified under EF2.

\emph{EF3} deals with market intermediation performed by broker-dealers, typically acting as market-makers for their clients. Lending platforms do handle funds for depositors, but not as a market-making intermediary (as would a decentralized exchange), so this EF is not applicable.

\emph{EF4} deals with the provision of credit enhancement to structured loan products. Because lending platforms do not provide credit enhancement, this EF is not applicable.

\emph{EF5} deals with entities creating asset- and mortgage-backed securities. Since lending platforms do not engage in this activity, this EF is not applicable.


\subsection{Stability risk factors}
\label{sec:leverage}
Recent literature has given leverage a central role in the buildup of asset price bubbles over time \cite{leverageCrisis, shadowBanking}. According to the ``marginal buyer of price" theory, market actors are heterogeneous in respect to their forecasts of asset prices. Leverage allows the most optimistic market actors to increase their asset purchases using borrowed funds, a situation which pushes asset prices up over time. When a negative price shock triggers margin calls, the most optimistic buyers are forced to sell their assets to avoid bankruptcy. These forced sales cause asset prices to decrease sharply. In essence, leverage increases volatility on both the up and down sides of asset price cycles.

This volatility is further increased when debt is chained across multiple financial products, as occurred with mortgage-backed securities prior to the financial crisis of 2007-09. The securitization of mortgages, while financially efficient, brought securities to the market that gave little information in regards to risk assumed. Because mortgage-backed securities were highly sensitive to house prices, a decline in those prices adversely affected security prices. The interlinkage of securities spread the adverse effect through these markets in a systemic way, leading to a run on multiple asset types \cite{gortonPanic}.

As mentioned in Section \ref{sec:activities}, the primary function of lending platforms is to facilitate debt creation. Following the marginal buyer theory, debt creation allows the most optimistic actors in the DeFi ecosystem to buy more assets. This leverage can both increase asset prices and decrease financial stability overall. Further, the interoperability of DeFi protocols allows highly leveraged actors to use their debt as collateral in other protocols, which further spreads risk across the system, creating an inherent risk of a system-wide run on assets.

Analyzing DFC allows us to gauge both the extent of debt financing, as well as its interconnectedness across multiple protocols, a critical step in assessing the risk present in the DeFi system. With blockchain transparent data, we can begin to estimate the flow of debt-backed funds temporally (over time) and cross-sectionally (across protocols).

According to our regulatory assessment, while leverage and interconnectedness may raise risk levels to the DeFi ecosystem, within which a negative shock within is unlikely to create significant effects in the broader financial system, considering the DeFi TVL. However, if DeFi protocols continue to grow in popularity, the risk to the global financial system, or portions thereof, will grow proportionally. Therefore, lending platforms are best described as an emerging risk which regulators may continue to monitor for future developments.

\section{Analysis methodology}
\label{sec:methodology}
\subsection{Scope of analysis}
Our goal is to identify the amount of DFC and assess the build-up of financial stability risks across the DeFi ecosystem, limiting our analysis across the following dimensions: time, protocols, and currencies.

First, we restrict our analysis to transactions between, and inclusive of, block 10,000,000 (May 4, 2020 01:22 PM UTC) and block 11,700,000 (January 21, 2021 04:50 PM UTC). From the beginning to end of our analysis period, the TVL of the DeFi ecosystem increased from \$0.9B to \$21.7B \cite{defiPulse}. 

Second, we choose four protocols to analyze: Aave, Compound, Maker, and Uniswap. Each project has been consistently ranked as one of the largest DeFi protocols in the past 12 months.  Maker, Aave, and Compound are (in order) the three largest DeFi protocols by TVL, as of April 30, 2021. Uniswap has been the largest decentralized exchange by volume and TVL, and consequently is involved in many exchanges that are intended to move funds between lending platforms \cite{defiPulse}.

Third, we filter transactions to include only those related to the currencies Bitcoin (BTC), Ether (ETH), Tether (USDT), USD Coin (USDC), and Dai (DAI)
covering 96\% of the amount borrowed in our selected lending platforms as of January 26, 2021, with protocol-level coverage ranging from 79\% to 100\% (see Table \ref{tab:coverage}). We also filter Uniswap transactions to include only exchanges between these five currencies.

\begin{table}[htbp]
    \caption{Borrowing by currency}
    \centering
    \begin{tabular}{@{} lccc @{}}
        \toprule
        \bf Currency    & \emph{Aave}   & \emph{Compound}   & \emph{Maker} \\
        \midrule
        WETH            &  64.5         &   97.3            & 0.0   \\
        WBTC            &  19.0         &   81.4            & 0.0    \\
        USDC            &  114.6        &   874.5           & 0.0  \\
        USDT            &  110.5        &   149.3           & 0.0     \\
        DAI             &  44.2         &   1,050.8         & 1,571.8     \\
        Other           & 93.3          &  63.5             & 0.0      \\
        \midrule
        Total           & 446.1         & 2,316.7           & 1,571.8 \\
        \midrule
        \% coverage     &  79\%         &  97\%             & 100\% \\
        \bottomrule \\
    \end{tabular}
    \linebreak \emph{In USD millions (point-in-time) as of Jan. 26, 2021}
    \linebreak \emph{Values from \cite{aaveStats} \cite{compStats} \cite{makerStats}}
    \label{tab:coverage}
\end{table}

\subsection{Heuristic selection}
In a transactional dataset, we may calculate the amount debt created or destroyed by identifying transactions that either withdraw or repay debt. However, we cannot identify debt being transferred from one protocol or wallet to another, as transactions are not specifically identified as debt-related or not. Therefore, we require a heuristic technique to estimate debt balances for each address. We consider an stylized scenario (Table \ref{tab:heuristic}) with three states ($S_0, S_1, S_2$) and two transactions ($\Delta_1, \Delta_2$). We assume a user holding currencies $\alpha$ and $\beta$, where the exchange rate between the currencies is 1. The user may hold funds in either wallet $W$ or lending platform $P$. At initial state $S_0$, the user holds in $W$ 100 non-debt units of $\beta$ and 100 debt-financed units of $\alpha$ (i.e. the user has withdrawn 100 $\alpha$ as debt in a separate lending transaction).

\begin{table*}[ht]
    \centering
    \caption{Heuristic outcomes}
    \setlength\tabcolsep{3pt} 
    
    \resizebox{\textwidth}{!}{%
        \subfloat[Heuristic 1: Proportional method]{
            \begin{tabular}{@{}lclccccc@{}}
                \toprule
                \bf Addr.               & \bf Curr.                 & \bf Type  & \bf $S_0$ & $\Delta_1$                    & \bf $S_1$
                    & $\Delta_2$                    & \bf $S_2$ \\ 
                \midrule
                \multirow{4}{*}{$W$}    & \multirow{2}{*}{$\alpha$} & Debt      & 100       & \textcolor{red}{-100}         & 0         
                    &                               & 0 \\
                                        &                           & Non-debt  & 0         &                               & 0      
                    &                               & 0 \\
                \cmidrule{3-8}
                                        & \multirow{2}{*}{$\beta$}  & Debt      & 0         & \textcolor{green}{+100} & 100    
                    & \textcolor{red}{-25}          & 75 \\
                                        &                           & Non-debt  & 100       &                               & 100        
                    & \textcolor{red}{-25}          & 75 \\
                \midrule
                \multirow{2}{*}{$P$}    & \multirow{2}{*}{$\beta$}  & Debt      & 0         &                               & 0     
                    & \textcolor{green}{+25}  & 25 \\
                                        &                           & Non-debt  & 0         &                               & 0      
                    & \textcolor{green}{+25}  & 25 \\
                \bottomrule
            \end{tabular}
        }
        \,
        \subfloat[Heuristic 2: First-out method]{
            \begin{tabular}{@{}lclccccc@{}}
                \toprule
                \bf Addr.               & \bf Curr.                 & \bf Type  & \bf $S_0$ & $\Delta_1$                    & \bf $S_1$
                    & $\Delta_2$                    & \bf $S_2$ \\ 
                \midrule
                \multirow{4}{*}{$W$}    & \multirow{2}{*}{$\alpha$} & Debt      & 100       & \textcolor{red}{-100}         & 0         
                    &                               & 0 \\
                                        &                           & Non-debt  & 0         &                               & 0      
                    &                               & 0 \\
                \cmidrule{3-8}
                                        & \multirow{2}{*}{$\beta$}  & Debt      & 0         & \textcolor{green}{+100} & 100    
                    & \textcolor{red}{-50}          & 50 \\
                                        &                           & Non-debt  & 100       &                               & 100        
                    &                               & 100 \\
                \midrule
                \multirow{2}{*}{$P$}    & \multirow{2}{*}{$\beta$}  & Debt      & 0         &                               & 0     
                    & \textcolor{green}{+50}  & 50 \\
                                        &                           & Non-debt  & 0         &                               & 0      
                    &                               & 0 \\
                \bottomrule
            \end{tabular}
        }
        \,
        \subfloat[Heuristic 3: Last-out method]{
            \begin{tabular}{@{}lclccccc@{}}
                \toprule
                \bf Addr.               & \bf Curr.                 & \bf Type  & \bf $S_0$ & $\Delta_1$                    & \bf $S_1$
                    & $\Delta_2$                    & \bf $S_2$ \\ 
                \midrule
                \multirow{4}{*}{$W$}    & \multirow{2}{*}{$\alpha$} & Debt      & 100       & \textcolor{red}{-100}         & 0         
                    &                               & 0 \\
                                        &                           & Non-debt  & 0         &                               & 0      
                    &                               & 0 \\
                \cmidrule{3-8}
                                        & \multirow{2}{*}{$\beta$}  & Debt      & 0         & \textcolor{green}{+100} & 100    
                    &                               & 100 \\
                                        &                           & Non-debt  & 100       &                               & 100        
                    & \textcolor{red}{-50}          & 50 \\
                \midrule
                \multirow{2}{*}{$P$}    & \multirow{2}{*}{$\beta$}  & Debt      & 0         &                               & 0     
                    &                               & 0 \\
                                        &                           & Non-debt  & 0         &                               & 0      
                    & \textcolor{green}{+50}  & 50 \\
                \bottomrule
            \end{tabular}
        }
    }
    \vspace{0.1em}
    
    \raggedright
    \emph{$S_0$}: Original state || 
    \emph{$\Delta_1$}: 100 $\alpha$ exchanged for 100 $\beta$ ||
    \emph{$\Delta_2$}: 50 $\beta$ deposited in Compound \\
    
    \label{tab:heuristic}
\end{table*}

In $\Delta_1$ the user exchanges 100 $\alpha$ for 100 $\beta$. Because the $\alpha$ balance is a debt balance, we also identify 100 $\beta$ as a debt balance. Therefore, at $S_1$, the user holds 200 $\beta$, 50\% of which are debt-financed units. 
In $\Delta_2$ the user deposits 50 units of $\beta$ into $P$. The user does not specifically identify what portion of the deposit, if any, is taken from the debt balance. In order to identify the debt-financed balances as of $S_2$, we consider three heuristics: proportional, first-out, and last-out. 
\begin{itemize}
    \item \emph{Proportional method}: Debt moves in proportion to the original debt percentage (50\%). Therefore, 25 units of the deposit are identified as debt.
    \item \emph{First-out method}: Debt-financed units are moved first. So, all 50 units of the deposit are identified as debt.
    \item \emph{Last-out method}: Debt-financed units are moved last. Therefore, 0 units of the deposit are identified as debt.
\end{itemize}

Both the proportional and last-out methods require a knowledge of the total balance of $\beta$ before the transaction occurs and would require gathering the balance of the relevant ERC-20 token for the calling address after every transaction. These balances can only be retrieved with access to an archive node, but their gathering would be resource-prohibitive.

The first-out method does not require a knowledge of the total ERC-20 balance, because the amount of debt moved is the maximum of the prior debt balance or the amount of the transaction. A downside of the method is that the calculated amount of debt involved in each transaction would likely be higher than the other two methods, because we assume the debt portion of the wallet balance always move first. The disparity would be most prevalent when the calculated debt balance is only a small percentage of the actual wallet balance.

Because of the technical limitations involved with using the proportional or last-out methods, we conclude that the first-out method is the most appropriate heuristic for our analysis.

\section{Algorithm specification}
\label{sec:algorithm}
\subsection{Log filtering}
Ethereum Logs $L$ are emitted when a named smart contract function is called and  
we analyze only logs created by smart contracts related to the Aave, Compound, Maker, or Uniswap protocols. 
We first define a set of relevant contract addresses $c$ for each protocol, and a set of function signatures $s$ used by each relevant contract. 
Then, we filter all Ethereum logs to find only the logs associated to contracts in $c$ and generated by calling functions in $s$. We define this set of logs as $E$.

\subsection{Address grouping}
While we could consider each individual address as a single user, this assumption would not accurately reflect user behavior on Ethereum, for two reasons: first, the open nature of the Ethereum blockchain allows individual users to generate multiple addresses, and second, protocols often create proxy contracts for each user interacting with the protocol. As a result, a single user could control multiple addresses or proxy contracts. Therefore, an analysis that assumes every user controls only one address would not accurately capture user behavior. Identifying users with multiple addresses is not a straightforward process, as all addresses are inherently pseudonymous. However, we can rely on the design architecture of the widely used DeFi protocols to perform a partial grouping of user-controlled addresses.

In the Maker protocol, each user must create a vault in order to deposit collateral and withdraw debt. A user may open an unlimited number of vaults, and the vault may be opened by the user's original address or by a proxy contract created specifically for the individual user.
If Address A creates Vault 1, Contract B creates Vault 2, and Address A created Contract B, then a single user must control both Vault 1 and Vault 2. All collateral and debt transactions from these vaults can be grouped together, and all transactions in other protocols by Address A and Contract B can also be grouped together. 

\begin{figure}
\caption{Address grouping and filtering}
\label{algo:grouping}

\begin{algorithmic}[0] 
\renewcommand{\algorithmicrequire}{\textbf{Input:}} \renewcommand{\algorithmicensure}{\textbf{Output:}}
\Require{Set of addresses $D$, $A$, $H$}
\Ensure{List of addresses $G_{ae}$}
\State $G_a, G_{ae} \leftarrow ()$ \;
\State $G_s \leftarrow \{\}$ \;

\For{$d \in D$} 
    \If{$d \cap G_s = \varnothing$} {Append $d$ to $G_a$ as new group}
    \Else{ Add $d_1, d_2, d_3$ to matching group in $G_a$ }
    \EndIf
    \State {$d \cup G_s$} \;
\EndFor

\For{$a \in A$} 
        \If{$a \cap G_s = \varnothing$} {Append $a$ to $G_a$ as new group}
        \Else{ Add $a$ to matching group in $G_a$}
        \EndIf
        \State $a \cup G_s$ \;
\EndFor
\For{$g \in G_a$}
    \State $\text{\emph{protocols\_in}} \leftarrow \sum\limits_{s\in A_{set}} y =
    \begin{cases}
        1    &\quad \text{if } g \subset s\\
        0    &\quad \text{else}
    \end{cases}$ \;
    \If{$\text{protocols\_in} > 1$} {Append $g$ to $G_{ae}$}
    \EndIf
\EndFor

\For {each pair $(r_1,r_2) \in H$}
    \For {each group $g \in G_{ae}$} 
        
    \If{ $r_1 \cap g \neq \varnothing$ and $r_2 \cap g = \varnothing$ }
        {$g \cup r_2$}
    \ElsIf{$r_1 \cap g \neq \varnothing$ and $r_2 \cap g_{next} \neq \varnothing$ }
        \State $merge( g, g_{next})$
        \Comment{$g_{next} \in G_{ae}$}
    \EndIf
    \EndFor
\EndFor

\end{algorithmic}
\end{figure}

This grouping logic could not be extended to the Aave and Compound lending platforms, as these protocols do not use a vault mechanism for borrowing. However, the thorough examination of smart contracts and their logs, allows us to infer heuristics based on the activity of users, which provides insights on users with multiple addresses. 

In the Aave protocol's smart contracts, the function \textit{repay}\footnote{ \url{https://docs.aave.com/developers/v/1.0/developing-on-aave/the-protocol/}}  allows users to repay a borrowed asset either partially or fully, for themselves or on behalf of a different user. The later indicates the same user repaying his/her own loan using funds from a different owned account. In the Uniswap protocol, the event \textit{Swap}\footnote{ \url{https://docs.uniswap.org/protocol/V2/reference/smart-contracts/pair}} allows us to identify cases where the sender and the destination addresses of token swaps is not the same, indicating a single user with two different EOA addresses. 
In the Compound's smart contracts, the function \textit{repayBorrowBehalf}\footnote{ \url{https://github.com/compound-finance/compound-protocol}} allows a sender to repay a borrow belonging to a borrower and in a similar direction we assume that the same user repays a borrow belonging to this single entity. These three defined heuristics return pairs of addresses, denoted as $H$ in the last part of our grouping algorithm. 

We define our grouping algorithm as follows: $A$ is the set of as all addresses initiating the transactions in $E$. While the address of the smart contract being called is stored in the log address field $L_a$, the address that initiates the call may be stored as one of the topics in $L_t$ or in the data field $L_d$, depending on the specifications of the contract function. Therefore, we search for the address in either $L_t$ or $L_d$. We notate this function-specific location as $L_x$.

We define $D$ as a set of all Maker vaults, with each item $d$ being a set $\{d_1,d_2,d_3\}$ containing the user address, the proxy address, and the urn handler address. Using the set of Maker vault addresses $D$, and the set of all addresses $A$, we then apply the algorithm shown in Algorithm \ref{algo:grouping}. From Algorithm \ref{algo:grouping}, we obtain a list of all addresses $G_a$, with related addresses grouped together (``address groups''). We then filter $G_a$ to contain only eligible address groups in which the associated addresses interact with two or more protocols ($G_{ae}$). We then identify more addresses from H, which belong to entities of eligible address groups and merge groups if one address is found in more than one groups.

\begin{figure}
\caption{Transaction classification}
\label{algo:classify}
\begin{algorithmic}[0]
\renewcommand{\algorithmicrequire}{\textbf{Input:}} \renewcommand{\algorithmicensure}{\textbf{Output:}}
\Require{List of addresses $G_{ae}$, set of logs $E$}
\Ensure{Sum of debt flows into each lending pool}
\For{$g \in G_{ae}$} 
    \State $E_g \leftarrow \{L \in E\mid L_x \subset g \}$ \;
    \For{$e \in E_g$}
        \If{$e$.trxType = Debt Create/Repay}
             \State walletDebt[$e$.currency] += $e$.amount \;
             
        \ElsIf{$e$.trxType=Col. Deposit/Withdraw}
          \State debtAmt = min($e$.amount, debt[$e$.currency]) \;
          \State walletDebt[$e$.currency] -= debtAmt \;
          \State platformDebt[$e$.currency] += debtAmt \;
          \If{$e$.trxType is Collateral Deposit} 
                \State sumDebtFlows += debtAmt 
          \EndIf
        
        \ElsIf{$e$.trxType = Swap}
            \State debtPct = min(1, debt[$e$.curr1] / $e$.sentAmt)
            \State walletDebt[$e$.curr1] -= $e$.sentAmt * debtPct
            \State walletDebt[$e$.curr2] += $e$.recvdAmt * debtPct
        \EndIf
    \EndFor
\EndFor
\end{algorithmic}
\end{figure}

\subsection{Transaction classification}

We then consider logs $E$, grouped by every $g$ in $G_{ae}$. We utilize Algorithm \ref{algo:classify}, which iterates through every transaction related to $g$. The algorithm calculates debt and non-debt balances for each address group, both point-in-time and cumulatively after every transaction. The purpose of the algorithm is to track debt balances across protocols, allowing for a quantification of debt withdrawn from one lending platform and deposited as collateral in another lending platform.

The classification of debt and non-debt balances are calculated according to the first-out heuristic previously described. All debt creation or repayment transactions are added to the address group's debt balance, which is disaggregated by currency. When swaps are performed, the percentage of debt in the currency being sent is used to calculate the debt added to the balance of the receiving currency. Collateral deposits and withdrawals are also recorded as debt based on the percentage of debt contained in the group's wallet at that point in time. For collateral deposits, which are the focal point of our analysis, we specifically track and aggregate the amount of debt involved in each transaction, for use in our final analysis.

\section{Data analysis}
\label{sec:analysis}
\subsection{Data collection}
All analysis was performed on an Ubuntu 20.04 virtual machine, and all records were stored in a MariaDB SQL database. Python, NodeJS, and R scripts were used for data retrieval and analysis.
For initial experiments, we filtered for the relevant data from the table `bigquery-public-data.\.crypto\_ethereum.logs' of Google BigQuery’s public Ethereum dataset, before setting up an Ethereum full Archive node (8-core Intel i7-11700 CPU, 4.8GHz, 32GB RAM, 10TB SSD). We collected transactions occurring between block 10,000,000 to 11,700,000, for protocols Aave, Compound, Maker, and Uniswap. Each protocol has upgraded the initial versions of their smart contracts, and in some cases multiple versions of the protocol may exist at one time. So, we collected data from the set of smart contracts associated to the protocol version with the most funds as of January 21, 2021. 
We took into consideration addresses belonging to exchanges in order to improve the effectiveness of our methods, using a public Kaggle dataset\footnote{ \url{https://www.kaggle.com/datasets/hamishhall/labelled-ethereum-addressesl}} containing almost 20,000 labeled addresses and a similar list of labeled addresses \cite{victor2020address}. We filtered the dataset to include transactions that met the following criteria: i) Involving BTC, ETH, USDT, USDC, DAI; ii) Collateral deposit/withdrawal, or debt creation/repayment (Aave, Compound, Maker); iii) Currency swaps (Uniswap).

We did not include liquidation transactions in our analysis, as we concluded that these transactions would not materially affect our analysis results. We also obtained pricing data for each of the five currencies analyzed: hourly pricing data for BTC, ETH, USDC, and DAI \cite{coinbaseAPI}, and daily pricing data for USDT \cite{nasdaqUSDT}. A summary of our dataset is shown in Table \ref{tab:summaryStats}.

\begin{table}[htb]
    \caption{Summary of dataset}
    \centering
    \resizebox{\columnwidth}{!}{%
        \begin{tabular}{@{} lcccc @{}}
            \toprule
            \textbf{Statistic}      & \textbf{Aave} & \textbf{Compound} & \textbf{Maker}    & \textbf{Uniswap} \\
            \midrule
            \multicolumn{5}{c}{\textbf{\emph{Counts}}} \\
            Unique addresses        & 22,645        &   293,929         &  10,157           &  210,214  \\
            Transactions            & 211,552       &   758,638         &  171,362          &  4,110,761  \\
            \midrule
            \multicolumn{5}{c}{\textbf{\emph{USD Totals}}} \\
            Collateral deposited    & 13,749        &   43,429          &  8,728            &  0  \\
            Collateral withdrawn    & 13,345        &   40,138          &  8,152            &  0  \\
            Debt created            & 3,252         &   17,934          &  4,512            &  0  \\
            Debt repaid             & 2,982         &   15,755          &  3,351            &  0  \\
            Currency swapped        & 0             &   0               &  0                &  26,653  \\
            \bottomrule \\
        \end{tabular}
    }
    \emph{USD values in millions, all values cumulative from period analyzed}
    \label{tab:summaryStats}
\end{table}

The majority of funds are channeled through the Compound protocol (collateral deposits and borrowing originations). The most common collateral deposited is WETH and the most common borrowed currency is DAI. Volume increased significantly across all protocols at the beginning of 2021.

\subsection{Algorithm results}
\begin{figure}[htb]
    \resizebox{\columnwidth}{!}{%
        \includegraphics{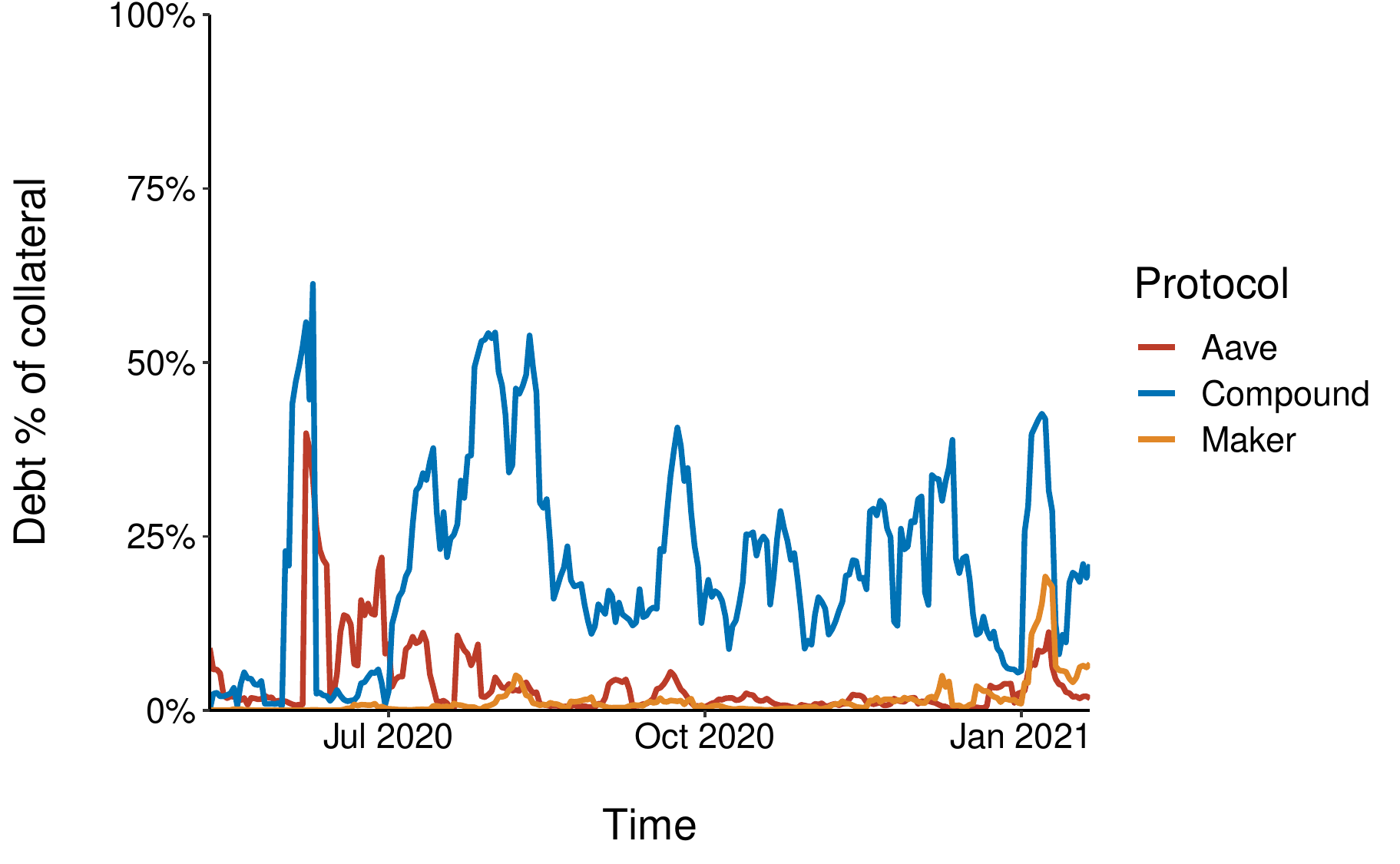}
    }
    \caption{Percentage of debt-financed collateral by protocol}
    \label{fig:debtPctProtocol}
\end{figure}

We applied Algorithm \ref{algo:grouping} to obtain address groupings, and Algorithm \ref{algo:classify} to quantify the amount of debt created and used in the DeFi ecosystem. Our results are set out in Table \ref{tab:algoResults}, and Figure \ref{fig:debtPctProtocol} shows additional detail by protocol. 
Our address grouping algorithm yielded 28,599 address groups that interacted with at least two protocols. These groups contained 47,944 unique addresses. The largest address group contained 97 addresses; however, only 0.12\% of groups ($n = 33$) had 10 or more addresses, and 45.1\% of groups ($n = 12,886$) had only one address.

We implemented and executed the self-authorization heuristic proposed in \cite{victor2020address} for Ethereum address clustering. ERC-20 token standard requires an approve function to allow another address to spend tokens on behalf of the actual owner. Upon successful function call, the event Approval is emitted, which contains the owner, spender and permitted amount. This functionality is exploited under the assumption that there are users that approve another address they own because of test purposes or risk distribution over several addresses with partial accessibility.

While this technique focuses on user activity only on Ethereum ERC-20 tokens exchange (not including DeFi protocols) and our algorithm focuses on interactions into DeFi protocols, the fact that 18 pairs of addresses belonging to same users were identified by both methods strengthens the validity of the approaches and showcases that clustering methods focusing on different Ethereum activities exhibit common results. 
As a future work ,we will compare the results with additional Ethereum heuristics such as \cite{victor2020address, beres2021blockchain, wu2022tutela }.

\begin{table}[htb]
    \caption{Summary of algorithm results}
    \centering
        \begin{tabular}{@{} lcccc @{}}
            \toprule
             & \multicolumn{3}{c}{\bf{Collateral deposited}} &  \\
            \cmidrule{2-4}
            \textbf{Month}      & \emph{Debt}   & \emph{Non-debt}   & \emph{Total}  & \textbf{Debt \%} \\
            \midrule
            May 2020            & 3             &   243             &  246          & 1.2  \\
            June 2020           & 86            &   2,326           &  2,412        & 3.6  \\
            July 2020           & 2,183         &   5,031           &  7,214        & 30.3  \\
            August 2020         & 1,001         &   4,511           &  5,511        & 18.2  \\
            September 2020      & 728           &   5,987           &  6,716        & 10.9  \\
            October 2020        & 589	        &   4,706           &  5,295        & 11.1  \\
            November 2020       & 1,277         &   7,778           &  9,054        & 14.1  \\
            December 2020       & 1,187	        &   8,956           &  10,143       & 11.7  \\
            January 2021        & 3,786         &   15,527          &  19,314       & 19.6  \\
            \bottomrule \\
            \multicolumn{5}{l}{\emph{All collateral values are monthly totals in USD millions}} \\
        \end{tabular}
    \label{tab:algoResults}
\end{table}

We observe that the percentage of DFC fluctuates significantly over the analysis period, from a low of 1\% in May 2020 to a high of 30\% in May 2020. From August 2020 through January 2021, the percentage of debt-financing fluctuated between 10 and 20\% of all collateral deposits in the five currencies. The protocol holding the largest amount of DFC is Compound. Although Maker vaults are used to create DAI, the Maker protocol itself does not hold significant percentages of DFC. 

For the majority of our analysis period, the currency most often deposited as DFC was DAI. However, beginning in late-2020, USDC became a significant source of DFC. WETH, USDT, and WBTC are not used in significant amounts as DFC.

\subsection{Correlation analysis}
We further analyze the correlation of the percentage of DFC to two variables: the changes in collateral deposited and in Ethereum price. If changes in DFC were correlated with a change in collateral deposited, this would suggest the presence of a feedback loop, in which increased collateral deposits lead to more DFC, which further increases collateral deposits, which leads to more DFC, and so on. The correlation of DFC with a change in Ethereum price would suggest that it is linked to speculation on the Ethereum price since it enables more borrowing. 
In our results (Table \ref{tab:correlation}), we observed a statistically significant correlation between an Ethereum price change in Day 1 and the percentage of DFC deposited in Day 2 ($p = 0.013$). That is, an increase in price was positively correlated with an increase in DFC the following day.

\begin{table}[htb]
    \caption{Variable correlation}
    \begin{tabular}{@{} lll @{}}
        \toprule
        \textbf{Variable 1} & \textbf{Variable 2}           & \textbf{$R$} \\
        \midrule
        Collateral change   & Debt percentage (next hour)   & -0.004 \\ 
        Collateral change   & Debt percentage (next day)    & 0.024 \\ 
        Price change        & Debt percentage (next hour)   & 0.016 \\ 
        Price change        & Debt percentage (next day)    & 0.154** \\ 
        \bottomrule \\
        \multicolumn{3}{l}{*$p<0.1$; **$p<0.05$; ***$p<0.01$} \\
        \multicolumn{3}{l}{\emph{$R=$ Pearson correlation coefficient}}
    \end{tabular}
    \vspace{-0.1cm}
    \label{tab:correlation}
\end{table}

Beyond observation of its existence, the relationship is not strong enough to draw definitive conclusions, particularly in regards to a casual relationship. An analysis with multiple independent variables would be required to isolate the effect price changes have on the prevalence of DFC.

\subsection{Interpretation of results}
\subsubsection{Leverage effects on stablecoins and protocols}

From our algorithm results, we conclude that the use of DFC is prevalent within the DeFi ecosystem, particularly with stablecoin collateral. In \cite{instability}, the risk of a ``deleveraging spiral'' breaking a stablecoin peg has already been noted. This risk is further sharpened based on the results of this paper, where it is clear that stablecoins are being used to create multiple layers of debt financing across protocols. 

DFC creates a high level of interconnectedness across lending platforms. This interconnectedness suggests that if one protocol suffers an operational shutdown or a run on deposits, the issue could quickly affect multiple DeFi protocols, which are linked through the use of DFC. In addition, it decreases the transparency of funding sources, as it is unclear whether collateral is fully owned by the depositor, or if it is actually debt owed to another protocol.

A salient step in future research may be the incorporation of DFC and its effects into a financial contagion model. Previous DeFi models (\cite{instability}), or models in the field of traditional banking networks (\cite{vulnerableBanks, contagionNetworks}), may be extended to predict how DFC affects the stability of the DeFi ecosystem overall.

\subsubsection{Analogies to traditional finance}
DFC can be considered be one form of ``yield farming,'' which is the process of moving currencies amongst protocols in the DeFi ecosystem, in order to capture the highest-yield investments \cite{yieldFarming}. Using DFC allows individuals to arbitrage interest rates across protocols, as well as to profit from expected price increases.

The use of DFC, and the phenomenon of yield farming more generally, has few corollaries to the traditional financial system, likely because interest rate differentials are not as easy to exploit as compared to DeFi. In academic research, previous financial literature has focused on the reuse of collateral, either through rehypothecation or securitization \cite{roleDebt, shadowBanking}. However, the use of debt as collateral has not been prevalent in traditional finance; therefore, the topic has been little explored in the literature.

One corollary to DFC may be the ``debt recycling'' process performed in Australia. Under Australian regulations, interest on loans used to make investments are tax-deductible, while mortgage interest is not. As a result, a common strategy is to draw on a home-equity line of credit, use the debt as collateral for an investment loan, and invest the funds in an income-producing asset \cite{ampDebtRecycle}. This cycle is similar to the process of creating DFC in DeFi. In the Australian regulatory environment, debt is recycled to take advantage of favorable tax rates; in the DeFi ecosystem, debt may be used to profit from diverging interest rates or expected price increases.

While debt recycling can be a cumbersome process to execute through a banking network, DFC is easier to accomplish through DeFi protocols because of the near-instantaneous settlement of transactions on the blockchain. If DeFi protocols continue to scale in popularity and usage, the open nature of DeFi may hold promise for other financial innovations to come.

\section{Conclusion}
\label{sec:conclusion}
In this paper, we first placed the activities of DeFi lending platforms into the FSB's NBFI regulatory framework. We concluded that lending platforms engaged in leverage creation and maturity transformation, and we classified lending platforms under the FSB's economic function ``Loan provision that is dependent on short-term funding.'' 

We then attempted to show the extent to which leverage from lending platforms has been used to create DFC. This use of DFC heightens the risk that negative price shocks will quickly spread through a chain of debt-backed lending positions, potentially feeding into a negative price spiral. We found that in some periods, 10-20\% of collateral deposits across five major currencies were debt-financed, and that USDC and DAI stablecoins were the currencies most commonly used as DFC, particularly in the Compound protocol. Based on these, market participants should be aware of the risk of adverse price shocks rapidly affecting the DeFi ecosystem.

Our quantitative insights were enabled by the use of a novel algorithm that grouped Ethereum accounts into user groups, and then classified debt balances across these user groups. Our analysis could be extended and improved through future work in two primary ways:
\begin{enumerate}
    \item Our ``first-out'' heuristic was chosen with time and technical limitations in mind. Additional analyses could undertake the resource-intensive task of estimating user balances at every point in time, giving a more accurate picture of user balances during a given period.
    \item Our account grouping algorithm relied on connecting accounts through four major DeFi applications. Additional grouping techniques could track activities across other DeFi protocols.
\end{enumerate}

Finally, we observe that the volume of capital flows into DeFi protocols is relatively small compared to the volume of flows into the broader universe of non-financial intermediaries. However, the popularity of DeFi has risen rapidly, and financial regulators would be well-served by monitoring this new form of credit intermediation more closely.

\bibliographystyle{IEEEtran}
\bibliography{references}

\vspace{12pt}
\end{document}